\documentclass{ws-procs975x65}            

\begin{document}
\title{Temperature calibration of the E and B experiment}

\author{Fran\c{c}ois Aubin$^*$, Asad M. Aboobaker, Chaoyun Bao, Christopher Geach, Shaul Hanany, Terry Jones, Jeff Klein, Michael Milligan, Kate Raach, Karl Young and Kyle Zilic}

\address{School of Physics and Astronomy, University of Minnesota,\\
Minneapolis, MN 55455, USA\\
$^*$E-mail: faubin@umn.edu}

\author{Kyle Helson, Andrei Korotkov, Valerie Marchenko and Gregory S. Tucker}
\address{Department of Physics, Brown University,\\
Providence, RI 02912, USA}

\author{Peter Ade and Enzo Pascale}
\address{School of Physics and Astronomy, Cardiff University,\\
Cardiff, CF24 3AA, United Kingdom}

\author{Derek Araujo, Daniel Chapman, Joy Didier, Seth Hillbrand, Bradley Johnson, Michele Limon, Amber D. Miller and Britt Reichborn-Kjennerud}
\address{Department of Physics, Columbia University,\\
New York, NY 10027, USA}

\author{Stephen Feeney  and Andrew Jaffe}
\address{Department of Physics, Imperial College,\\
London, SW7 2AZ, United Kingdom}

\author{Radek Stompor}
\address{Laboratoire Astroparticule et Cosmologie (APC),\\
Paris, 75205, France}

\author{Matthieu Tristram}
\address{Laboratoire de l'Acc\'el\'erateur Lin\'eaire, Universit\'e Paris-Sud,\\
Orsay, 91898, France}

\author{Matt Dobbs, Kevin Macdermid and Graeme Smecher}
\address{Department of Physics, McGill University,\\
Montr\'eal, H3A 2T8 Qu\'ebec, Canada}

\author{Julian Borrill and Theodore Kisner}
\address{National Energy Research Supercomputing Center, Lawrence Berkeley National Laboratory,\\
Berkeley, CA 94720, USA}

\author{Gene Hilton, Johannes Hubmayr and Carl Reintsema}
\address{National Institute of Standards and Technology,\\
Boulder, CO 80305, USA}
\newpage

\author{Carlo Baccigalupi and Giuseppe Puglisi}
\address{Astrophysics Sector, SISSA,\\
Trieste, 34014, Italy}

\author{Adrian Lee and Ben Westbrook}
\address{Department of Physics, University of California, Berkeley,\\
Berkeley, CA 94720, USA}

\author{Lorne Levinson and Ilan Sagiv}
\address{Faculty of Physics, Weizmann Institute of Science,\\
Rehovot, 76100, Israel}

\begin{abstract}
The E and B Experiment (EBEX) is a balloon-borne polarimeter designed to measure the polarization of the cosmic microwave background radiation and to characterize the polarization of galactic dust.
EBEX was launched December 29, 2012 and circumnavigated Antarctica observing $\sim$6,000~square degrees of sky during 11~days at three frequency bands centered around 150, 250 and 410 GHz.
EBEX was the first experiment to operate a kilo-pixel array of transition-edge sensor bolometers and a continuously rotating achromatic half-wave plate aboard a balloon platform. It also pioneered the use of detector readout based on digital frequency domain multiplexing.

We describe the temperature calibration of the experiment.
The gain response of the experiment is calibrated using a two-step iterative process.
We use signals measured on passes across the Galactic plane to convert from readout-system counts to power.
The effective smoothing scale of the EBEX optics and the star camera-to-detector offset angles are determined through $\chi^2$ minimization using the compact HII region RCW 38.
This two-step process is initially performed with parameters measured before the EBEX 2013 flight and then repeated until the calibration factor and parameters converge.
\end{abstract}


\bodymatter

\section{Introduction}

The E and B Experiment (EBEX) was a long duration balloon-borne cosmic microwave background (CMB) polarimeter.\cite{2010SPIE.7741E..37R}
EBEX observed the sky at three frequency bands centered on 150, 250 and 410 GHz, to enable foreground characterization, which is essential for primordial gravity wave detection.\cite{2015PhRvL.114j1301B}
The EBEX telescope
was an off-axis Gregorian Mizuguchi-Dragone telescope with a 1.5 m primary mirror.
Two redundant star cameras provided discrete pointing solutions with arcsecond resolution.
The pointing between star camera solutions was interpolated using two redundant three-axis gyroscopes.\cite{7119010}
The polarimetry was achieved through a combination of a polarizing grid and a continuously rotating achromatic half-wave plate (AHWP) levitated above high temperature superconducting bearings.\cite{2011SPIE.8150E..04K}
The AHWP was composed of a five-stack of sapphire half-wave plates.
It was maintained at 4~K  and located near the aperture stop of the optical system.
EBEX had two focal planes containing spider-web transition edge sensor (TES) bolometers operated with a bath temperature of 250~mK. 
The bolometers were fabricated with low thermal conductance as appropriate for the loading expected at balloon flight altitudes.\cite{2012JLTP..167..885W}
They were readout with superconducting quantum interference device (SQUID) pre-amplifiers, and using a digital frequency domain multiplexing (DfMUX) system with multiplexing factor of 16.\cite{2008ITNS...55...21D}

\begin{figure}[h]
  \begin{center}
    \includegraphics[height=1.1in]{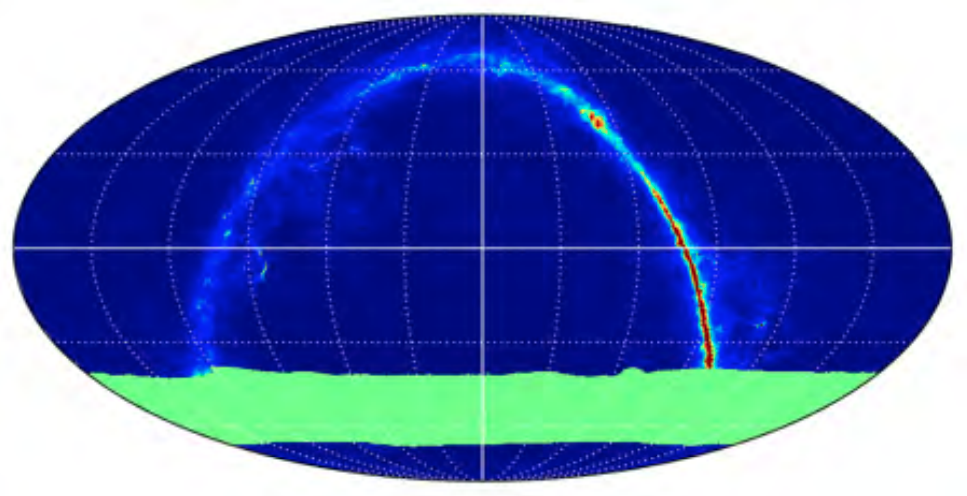}
    \caption{The $\sim$6,000 square degrees of sky observed during the EBEX 2013 flight (green) in equatorial coordinates.
    }
    \label{fig:ebex}
  \end{center}
\end{figure}

\section{The EBEX 2013 Flight}

EBEX was launched on December 29, 2012 from Antarctica and maintained an altitude of approximately 35~km for 25.5~days.
The cryogens lasted 11~days as originally designed for a typical single circumnavigation flight around Antarctica.
When the payload reached float altitude we successfully tuned 955 TES bolometers.
Overheating of the azimuth motor controller rendered azimuth control inoperable. 
We therefore implemented a revised scan pattern. We maintained a constant elevation of 54~degrees and the azimuthal motion consisted of a superposition of free rotations and torsional oscillations at the natural frequency of the flight-line.
The gondola scan speed was below 1$^o$/s for more than 99\% of the flight.
Fig.~\ref{fig:ebex} shows the resulting coverage of $\sim$6,000 square degrees of sky between -69$^o$ and -38$^o$ DEC.

The AHWP was set to continuous 1.235~Hz rotation prior to launch.
It continued to rotate through launch and throughout the entire flight with the exception of several short periods during which we conducted rotationless calibration observations.
The total number of rotations at float was 651,000.

The EBEX 2013 flight marks the first balloon-borne operation of a monolithic array of kilopixel TES bolometers, the first implementation of a frequency multiplexing factor of 16 on any telescope, the first implementation of a 4~K continuously rotating AHWP, and the first use of a 4~K superconducting magnetic bearing for astrophysical applications.\cite{2014SPIE.9153E..11M}

\section{Temperature Calibration}

To make maps of calibration sources we apply the following steps sequentially to the time ordered data: fitting and removal of an AHWP rotation synchronous signal, deglitching, flagging for bad data, band-pass filtering between 0.1 and 8.0~Hz, downsampling and noise-weighted binning in sky coordinates.
Data recorded with a scan speed such that the calibration source signal is below the 0.1~Hz high-pass are not included.
We quantify the detector noise by fitting a gaussian function to the time-ordered data distribution for each source crossing.
We do not deconvolve the temporal response of the bolometer.
The predicted source stretching given a scan speed of 1$^o$/s and bolometer time constant faster than 20~ms is less than 12\%.
We do not observe source stretching in jackknife maps made with azimuth scans in opposing direction.\cite{1998MNRAS.299..653H}
At this stage the maps have units of readout-system counts. 

The calibration consists of fitting the source map from each detector to a detector-specific processed reference map using a single multiplicative 'calibration factor'.
The processed reference maps are made in the following way.
We generate a temperature map of the calibration source by summing the Planck component maps, which have been scaled to and integrated over each of the three EBEX frequency bands.\cite{2015arXiv150201588P}
Using the EBEX pointing information we create detector-specific time-ordered data by scanning the temperature map.
These data are subject to the same previously described map-making processing as the original EBEX data.
The processed reference map has units of power (or equivalent CMB temperature) and the calibration factor therefore has units of power/count.
We quantify the quality of each of the detector-specific calibration factor by calculating the correlation coefficient between the processed reference and the calibrated source maps.

The temperature calibration of the EBEX detectors is challenging because of an irregular coverage of the calibrator sources. 
The primary calibration source is the Galactic plane, which is the only source common to all detectors.
We do not calibrate to the CMB due to its lower signal-to-noise ratio for individual detector maps.
RCW 38, which is a compact, quite intense, HII region in the Galactic plane, was scanned sporadically by only a subset of the detectors.
The detector calibration requires simultaneous optimization of the following 'calibration parameters': the effective smoothing scale of the telescope and the star camera-to-detector offset angles.
A calibration factor is first produced for every detector using the best known calibration parameters.
The calibration parameters are then optimized using the latest calibration factors and the scans of RCW 38.
This two-step process is initially performed with the calibration parameters measured before the flight and then repeated with the updated parameters until the calibration factor and its parameters converge.

\def\figsubcap#1{\par\noindent\centering\footnotesize(#1)}
\begin{figure}[h]
  \begin{center}
    \parbox{1.4in}{\includegraphics[height=1.3in]{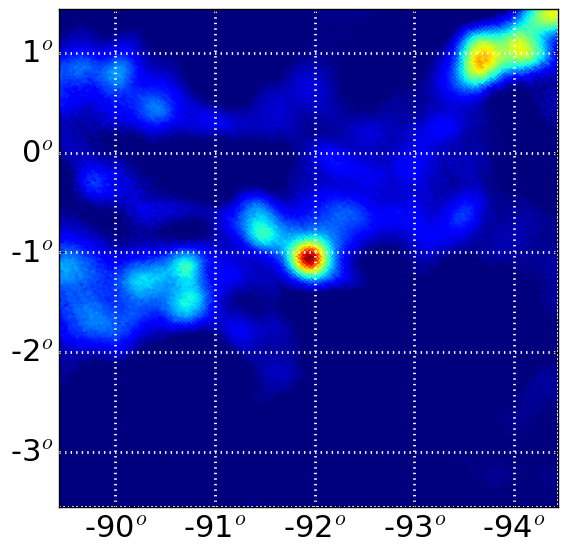}\figsubcap{a}}\hspace*{4pt}
    \parbox{1.4in}{\includegraphics[height=1.3in]{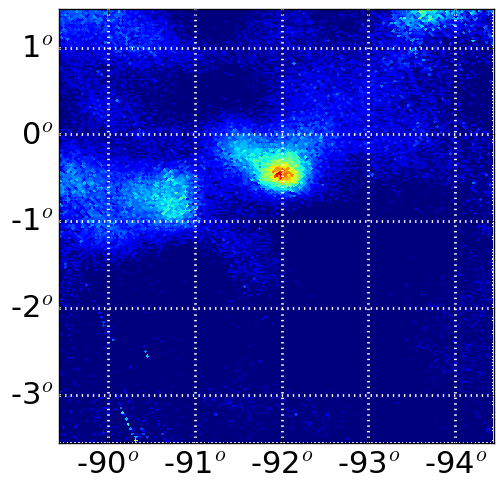}\figsubcap{b}}\hspace*{4pt}
    \parbox{1.4in}{\includegraphics[height=1.3in]{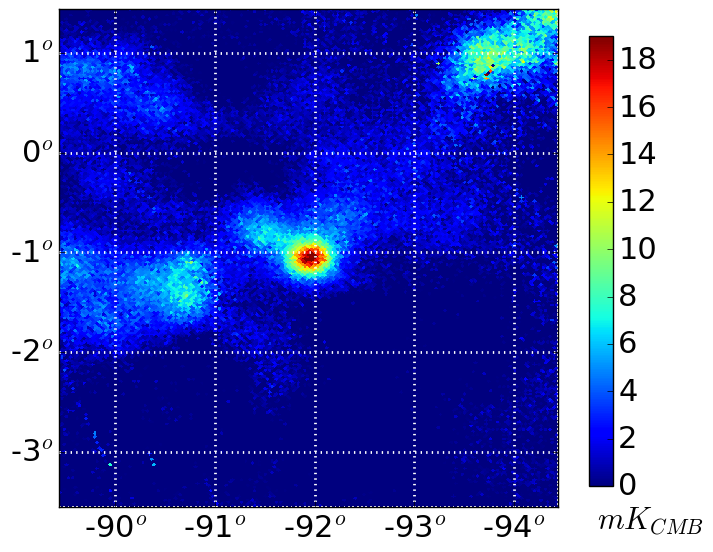}\figsubcap{c}}
    \caption{The HII region RCW 38 for 227 of the 250~GHz detectors in galactic coordinates.
      (a) Processed reference map smoothed to 15.0 arcminutes.
      (b) EBEX map with pre-flight star camera-to-detector offset angles.
      (c) EBEX map with optimized star camera-to-detector offset angles.}
    \label{fig:ebexRCW38Map}
  \end{center}
\end{figure}

We make several simplifying assumption during the iterative optimization process.
We model the boresight-to-detector offset angles using the known relative positions of horns on the monolithic focal plane and plate scale calculations using the nominal telescope design.
We assume that the star camera-to-boresight offset angles are the same for all detectors within each of the 14 bolometer wafers.
We assume that the effective smoothing scale is the same for all detectors at a given frequency band.
These assumptions and the decision to not deconvolve the temporal response function of the detector reduce the number of calibration parameters from 5,730 to 1,000.
The maps from all detectors on the same wafer and in the same frequency band are co-added during the star camera-to-boresight offset angle and the effective smoothing scale optimization, respectively.

\begin{figure}
  \begin{center}
    \parbox{4.2in}{\includegraphics[width=4.1in]{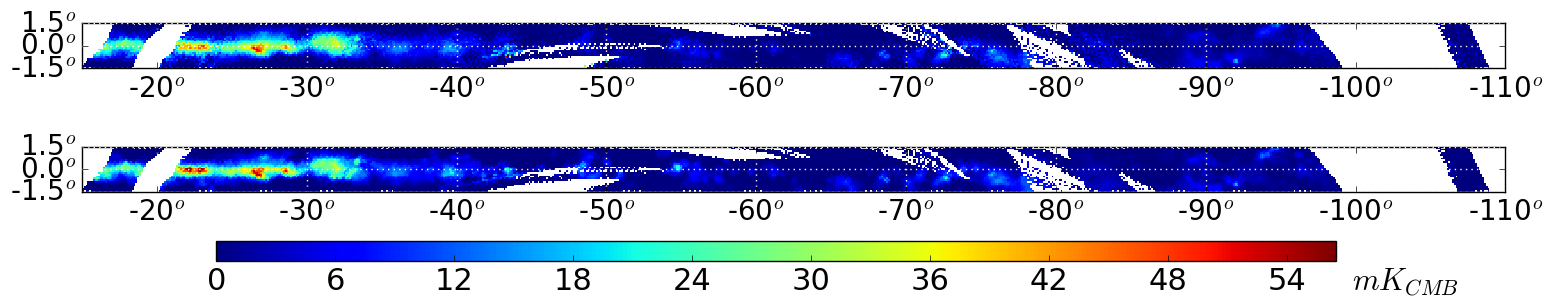}\figsubcap{a}}\\
    \parbox{1.7in}{\includegraphics[width=1.6in]{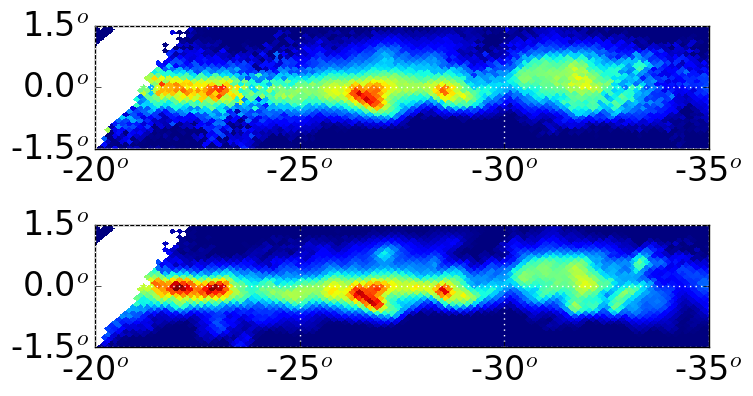}\figsubcap{b}}\hspace*{4pt}
    \parbox{1.7in}{\includegraphics[width=1.6in]{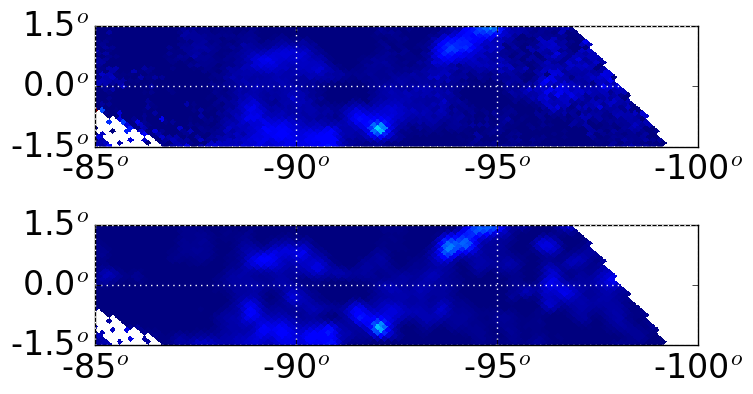}\figsubcap{c}}
    \caption{(a) The Galactic plane in the 250 GHz EBEX frequency band.
      Top panel: the optimized EBEX temperature map using 236 detectors.
      Bottom panel: the corresponding processed reference map smoothed to 15.0 arcminutes.
      (b) \& (c) Zoom on two different areas from (a).}
    \label{fig:ebexGalMap}
  \end{center}
\end{figure}

EBEX maps produced with different star camera-to-boresight offset angles are compared to processed reference maps that are smoothed to various scales.
The optimal calibration parameters are chosen by minimizing the $\chi^2$ between the processed reference and the EBEX maps.
Fig.~\ref{fig:ebexRCW38Map} shows the processed reference map of RCW 38 with an effective smoothing scale of 15.0 arcminutes and the EBEX maps of RCW 38 with pre-flight and optimized star camera-to-boresight offset angles for 227 of the 250~GHz detectors.
Only detectors which have a calibration factor with a correlation coefficient greater than 0.3 are included in the comparison.
Moreover,  detectors which have a calibration factor that is more than 5$\sigma$ away from the center of the distribution of all calibration factors within the same frequency band are rejected. 
Fig.~\ref{fig:ebexGalMap} shows the co-added map of the Galactic plane for 236 of the 250~GHz detectors and the corresponding processed reference map.
This iterative analysis has also been processed for the 150 and 410~GHz detectors and shows similar results.

\section{Conclusion}

We discussed the temperature calibration of the EBEX 2013 flight.
The pipeline is based on scans of the Galactic plane and RCW 38.
It is complete and an initial calibration has been produced for all detectors.

\section*{Acknowledgements}

EBEX is a NASA supported mission through grants number NNX08AG40G and NNX07AP36H.
We thank Columbia Scientific Balloon Facility for their enthusiastic support of EBEX. 
The authors acknowledge the Minnesota Supercomputing Institute (MSI) at the University of Minnesota for providing resources that contributed to the research results reported within this paper.
The research described in this paper used facilities of the Midwest Nano Infrastructure Corridor (MINIC), a part of the National Nanotechnology Coordinated Infrastructure (NNCI) program of the National Science Foundation.
This research used resources of the National Energy Research Scientific Computing Center, which is supported by the office of Science of the U.S. Department of Energy under contract No. DE-AC02-05CH11231.
The McGill authors acknowledge funding from the Canadian Space Agency, Natural Sciences and Engineering Research Council, Canadian Institute for Advanced Research, Canadian Foundation for Innovation and Canada Research Chairs program.

\bibliographystyle{ws-procs975x65}
\bibliography{ws-pro-sample}

\end{document}